# Programmable Generation of Terahertz Bursts in Chirped-Pulse Laser Amplification

Vinzenz Stummer,[1] Tobias Flöry,[1,2] Gergő Krizsán,[3] Gyula Polónyi[3,4] Edgar Kaksis[1], Audrius Pugžlys[1,5], József András Fülöp[3,4,6], Andrius Baltuška[1,5,*]

[1]*Photonics Institute, TU Wien, Gusshausstraße 27-387, A-1040 Vienna, Austria*
[2]*Institute of Theoretical Chemistry, University of Vienna, Währingerstraße 17, A-1090 Vienna, Austria*
[3]*Szentágothai Research Centre, University of Pécs, Ifjúság ú. 20, H-7624 Pécs, Hungary*
[4]*MTA-PTE High-Field Terahertz Research Group, Ifjúság ú. 6, H-7624 Pécs, Hungary*
[5]*Center for Physical Sciences & Technology, Savanoriu Ave. 231, LT-02300 Vilnius, Lithuania*
[6]*ELI-ALPS, ELI-HU Nonprofit Ltd., Wolfgang Sandner u. 3, H-6728 Szeged, Hungary*
*\*andrius.baltuska@tuwien.ac.at*

**Abstract:** Amplified bursts of laser pulses are sought for various machining[1], deposition[2], spectroscopic[3] and strong-field applications[4]. Standard frequency- and time-domain techniques for pulse division[5-14] become inadequate when intraburst repetition rates reach the terahertz (THz) range as a consequence of inaccessible spectral resolution, requirement for interferometric stability, and collapse of the chirped pulse amplification (CPA) concept due to the loss of usable bandwidth needed for safe temporal stretching. Avoiding the burst amplification challenge and resorting to a lossy post-division of an isolated laser pulse after CPA leaves the limitations of frequency- and time-domain techniques unsolved. In this letter, we demonstrate an approach that successfully combines amplitude and phase shaping of THz bursts, formed using the Vernier effect, with active stabilization of spectral modes and efficient energy extraction from a CPA regenerative amplifier. As proof of concept, the amplified bursts of femtosecond near-infrared pulses are down-converted into tunable THz-frequency pulses via optical rectification.

Laser pulse bursts with high energies up to multi-millijoules at various intraburst repetition rates have already found their way into a number of applications, such as materials processing[1], pulsed laser deposition[2], laser induced breakdown spectroscopy (LIBS)[3] and seeding of free electron lasers[4]. However, for high intraburst repetition rates, there exists a fundamental limitation of the achievable burst energy in conventional CPA systems which prevents access to the multi-mJ regime. This limitation originates from the buildup of spectral modes that are mapped into the time domain as strong spikes on the temporal intensity profile of the stretched overlapping pulses circulating in the laser amplifier. The motivation of this work is to propose and demonstrate an approach which allows suppression of spectral mode formation by programming phases of individual pulses at THz intraburst repetition rates directly in the time domain.

Time-domain methods relying on electro-optic or acousto-optic modulators[5-7] are by far too slow to reach the THz regime. By using Fourier-synthesis methods or direct space-to-time conversion[8], a pulse can be shaped with masks, programmable spatial light modulators[9], or arrayed-waveguide gratings[10-13]. However, this approach is inadequate for versatile burst formation by broadband pulse shaping at THz modulation frequencies due to an impossibly high number of spectral channels needed to satisfy both the spectral bandwidth and resolution criteria. Consequently, the programmable time spacing for femtosecond pulses is restricted to several picoseconds, making entire classes of experiments, such as studies of a rotational molecular response, inaccessible to existing spatial light modulators. Interferometric pulse division and re-stacking methods can provide THz modulation frequencies but their flexibility and programmability is strongly restricted[14].

In this work, the Vernier effect[15] is utilized in a burst generation method, which is capable of providing arbitrarily short temporal separations $\Delta t$ of pulses as determined by the difference of the roundtrip time $t_{MO}$ of a master oscillator (MO) and the roundtrip time $t_{RA}$ of a regenerative amplifier (RA). The enormous potential of the Vernier effect is explained by the following advantages: 1. By adjusting the difference in cavity length, the pulse separation can be tuned continuously down to within the duration of a single compressed pulse. 2. The number of pulses per burst can be chosen arbitrarily, as the upper limit on the burst duration is only limited by the time of cavity roundtrip and Pockels cell switching. 3. Intraburst pulse shaping can be decoupled from the burst formation process by shaping pulses before injection into the RA. This enables direct time-domain amplitude and phase control at the THz intraburst repetition rate using slow modulators operating at the MO repetition rate.

Burst-mode amplification is frequently perceived as more efficient compared to single-pulse amplification because the stored laser energy can be divided up among $N$ pulses that offer an $N$-fold increase of the effective pulse duration in the amplifier and a $\sqrt{N}$-times energy scaling. This advantage is explored in many variants of divided-pulse amplification[16-19]. However, no such advantage exists for pulse separations that are significantly shorter than the stretched pulse duration in conventional CPA. On the contrary, as illustrated in Fig. 1(a), the energy safely extractable from a chirped pulse amplifier rapidly drops as $N$ increases. A burst composed of $N$ spectrally overlapping pulses with constant phase offsets exhibits discrete spectral lines, spaced at the intraburst repetition rate, as an outcome of spectral interference. Therefore, compared to single-pulse operation, the peak spectral intensities inside the amplifier dramatically increase by a factor of $N^2$ for the same output energy. This $N^2$ scaling reflects the growth of temporal peak intensities, as a result of frequency-to-time mapping at high chirp rates. To resolve this detrimental consequence of mode formation, we employ a phase-scrambling technique which inhibits constructive interference by applying programmable phase offsets $\Phi_{offset,n}$ to each individual intraburst pulse. These offsets are combined with a constant pulse-to-pulse phase slip $\Delta\Phi_{slip} = k\Delta L = k(L_{MO}-L_{RA})$ which follows from the mismatch of the MO-RA cavity roundtrip lengths $\Delta L$, with $k$ being the wavenumber. The phase differences between subsequent intraburst pulses become

$$\Phi_n - \Phi_{n-1} = \Delta\Phi_{slip} + (\Phi_{offset,n} - \Phi_{offset,n-1}) \quad (n = 2, 3, ..., N). \quad (1)$$

By an appropriate choice of the phase offsets, the burst spectrum is shaped to maximally resemble the single-pulse spectrum. The measured burst spectra can be well reproduced by numerical calculations based on the measured single-pulse spectrum and the intraburst pulse-to-pulse phase slip as the only fit-parameter (Fig. 1(b,c)).

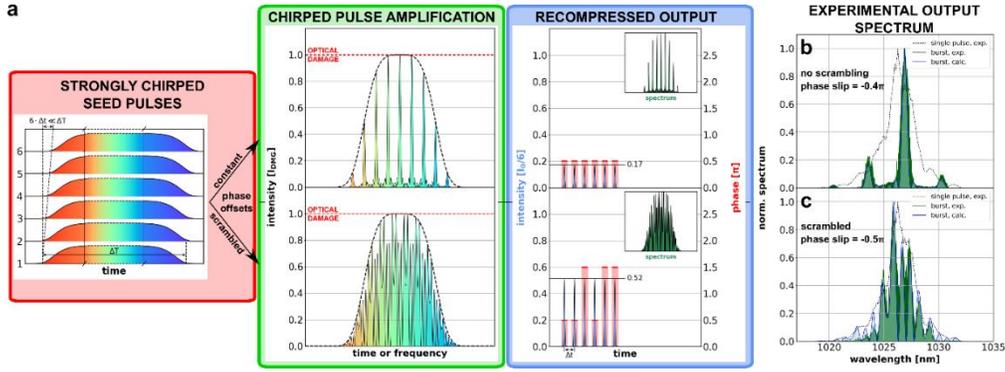

**Fig. 1. Safe chirped amplification of a burst utilizing phase-scrambling.** a, left) Composition of the seed burst from strongly chirped overlapping intraburst pulses. a, middle) Spectral or temporal intensity of the amplified strongly chirped burst, normalized to the optical damage threshold intensity $I_{DMG}$ of the RA, with pronounced spectral modes in the case of constant phase offsets (upper panel). This can be avoided by programmable scrambling of the phase offsets (lower panel). An amplified single chirped pulse is shown for comparison (black dashed line). a, right) Time-domain representation of intensity (blue) and phase (red) after compression with the intensity normalized to $I_0/N$, with $I_0$ being the maximum output peak intensity of a single pulse and N the pulse number. The inset shows the spectrum (green). b,c) Experimental results showing the single-pulse spectrum (black dashed line) and the burst spectra for N=6 (green) with the programmable phase offset being b) constant, c) scrambled with $\pi$ phase-shifts. The calculated spectra are also shown (blue lines).

We explored the potential of our approach by determining the maximum extractable energy achievable by phase-scrambling as a function of the burst pulse number, and benchmarked it against the single-pulse case (Fig. 2). We determined optimized sets of phase offsets in two ways: first, by a numerical optimization algorithm (L-BFGS-B)[20] and second, by calculating the global maximum energy with restricting the phase offset to 0 and $\pi$ ($\pi$-shift). Without phase-scrambling, the maximum extractable energy is inversely proportional to the pulse number. In contrast, for both illustrated phase-scrambling cases, the relative extractable energy stays reliably above 50% and approaches approximately 70% of the maximum single-pulse energy for higher pulse numbers.

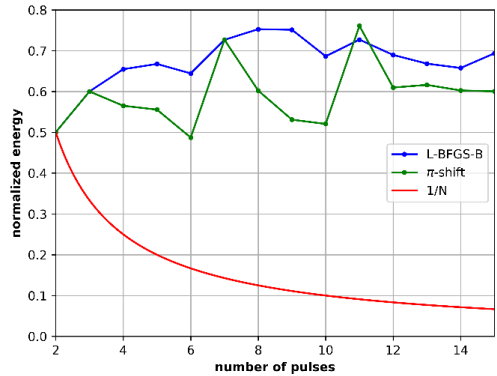

**Fig.2. Maximum extractable energy without exceeding the optical damage threshold in burst-mode normalized to the single-pulse case vs. pulse number N.** Calculation with a programmable phase offset which is constant (red), optimized by a numerical L-BFGS-B algorithm (blue) and the global maximum when allowing only 0 or $\pi$ phase offsets (green).

Our Vernier RA setup with intraburst phase slip stabilization is presented in Fig. 3. A periodic pulse train is generated from an Yb:KGW MO with $f_{MO} = 76$ MHz repetition rate and the pulses are stretched to approximately 150 picoseconds. For the control of individual pulse amplitudes and phases an acousto-optic modulator (AOM) is used. The RA Pockels cell (PC) operates synchronously with the AOM to accumulate the AOM-diffracted seed pulses in the RA cavity. The function of the PC is modified compared to conventional on/off-operation by allowing to switch to an intermediate PC voltage level for pulse accumulation in the cavity, enabling subsequent seed pulses to enter the cavity while only partially coupling out already accumulated pulses. The PC voltage is set such that the induced losses are compensated by the round-trip gain, resulting in uniform burst formation. After the desired number of burst pulses is reached, the PC voltage is raised further to lock and amplify the burst in the cavity. In order to reach multi-millijoule energies, the burst is amplified further in a cryogenically-cooled Yb:CaF$_2$ booster RA before it is recompressed to 250 fs by a grating compressor.

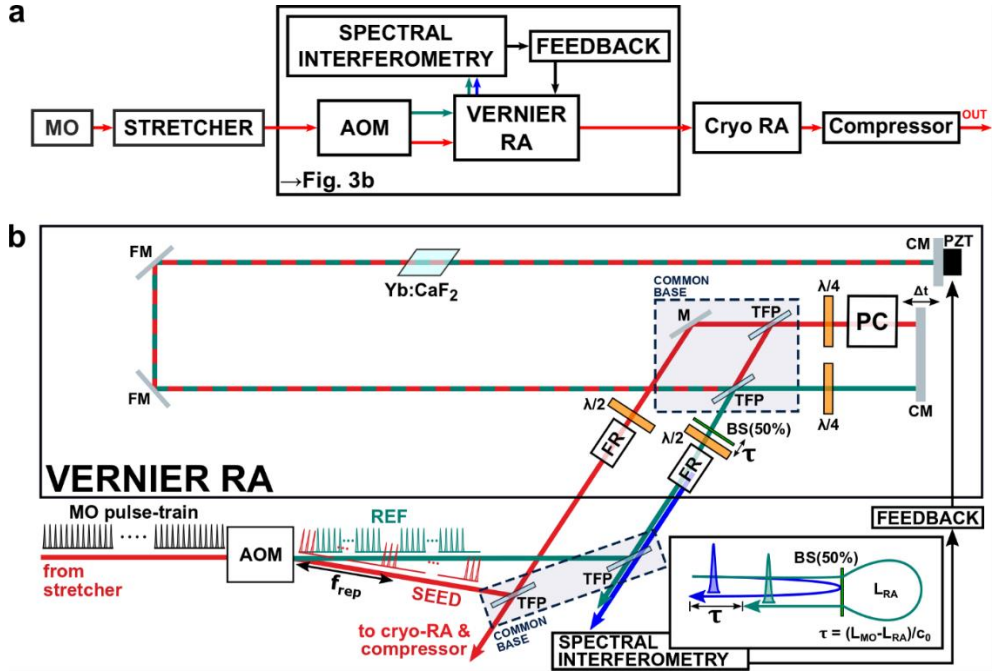

**Fig. 3: Vernier setup.** a) Block diagram of the complete experimental setup comprising a master oscillator (Light Conversion Flint), two regenerative amplifiers (one as preamplifier, one as booster amplifier), a stretcher, a compressor and an acousto-optical modulator for separating seed and reference pulses. The regenerative amplifiers are home-made table-top systems, both with Yb:CaF$_2$ crystals as active elements, where for the booster amplifier the crystal was cryo-cooled to approximately 100 K. b) More detailed scheme of the Vernier RA with burst formation and phase slip stabilization. Pulses with a sufficiently long temporal spacing are diffracted from the MO pulse train (black) with an AOM, which enables individual amplitude and phase modulation. The diffracted pulses (red) are used as burst seed and the non-diffracted pulses (turquoise) are used as high-repetition-rate reference for measuring the phase slip drift. The seed pulses are accumulated and amplified in the RA cavity. Inset: Concept of the interferometric phase slip drift measurement. The reference pulses are split into two pulses. One (blue) is reflected from a beamsplitter (BS) and another (turquoise) fulfills a roundtrip in the RA cavity. The timing is set such that the transmitted part of a pulse spectrally interferes with the reflection of the next pulse.

Thermal drift and mechanical vibrations can lead to variations in the cavity roundtrip times thus affecting the intraburst phase slip and the burst spectrum. Therefore, an active feedback control of the cavity length difference was implemented to compensate such influences (Fig. 4), without the need of an external reference cavity. This method can be seen as an intraburst counterpart of widely established CEP stabilization techniques[21-23] that, to date, have been applied to continuous pulse trains of mode-locked oscillators. The differential cavity length drift is stabilized by monitoring the spectral interference of consecutive pulses in the non-diffracted beam at the MO repetition rate $f_{MO}$ (Fig. 3). One of the interfering reference pulses is reflected off a beamsplitter and the other one undergoes one complete RA roundtrip. The active feedback control is realized as a software-implemented PID controller with a piezoelectric transducer (PZT) attached to one of the mirrors in the RA cavity and by evaluating the reference interference spectrum.

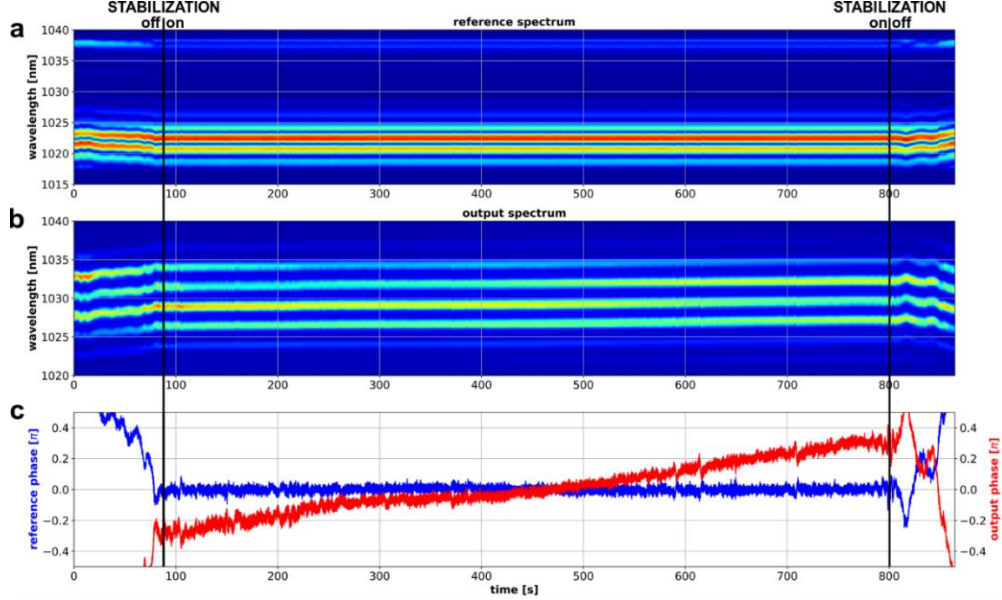

**Fig. 4: Intraburst phase slip stabilization results.** Spectrum over time of **a)** an interferometric reference (in-loop) **b)** the output burst (N=4, Δt=3ps, out-of-loop). **c)** Phase of the interferometric reference (blue) and the intraburst phase slip (red) over time. The intraburst phase slip stabilization is realized by controlling the cavity length difference with an active feedback loop by applying an interferometric in-loop reference. The reference interference spectrum shows a mitigated bandwidth, because of spectral shaping of the seed pulses for precompensation of gain-narrowing. The stabilization reduces strongly the phase slip drift and thus leads to a stable burst spectrum over an extended time-period.

As to demonstrate control over THz intraburst repetition rates, we perform generation of THz-pulse bursts by optical rectification in $LiNbO_3$ using tilted-pulse-front pumping (TPFP)[24,25]. SHG autocorrelations of the NIR burst driver and recorded linear autocorrelations of generated THz transients using a Michelson interferometer together with spectra retrieved by Fourier transformation can be seen in Fig. 5. The measurements were performed for variable pulse spacing with fixed pulse number (Fig. 5(a), $N=6$) and for variable pulse number with fixed pulse spacing (Fig. 5(b), $\Delta t=3$ ps). As expected, the continuously-tunable intraburst repetition rate is determined by the inverse pulse spacing $1/\Delta t$, translating into the lowest-order THz frequency, whereas the bandwidth $\Delta f$ of the generated THz peak scales inversely to the product of pulse number and spacing $1/(N \cdot \Delta t)$, with values $\Delta f_{N=2}$=138 GHz, $\Delta f_{N=4}$=58 GHz, $\Delta f_{N=6}$=50 GHz. The spectra exhibit a high spectral contrast that progessively improves with $N$. Higher-harmonics of the THz-signal are observed as well, in accordance with the Fourier transformation of a windowed pulse train. The THz pulse source demonstrated here offers a unique combination of versatile pulse shaping capability and scalability to high energies, thereby surpassing many other methods[26,27].

In conclusion, we demonstrate chirped-pulse amplification of femtosecond pulse bursts with THz intraburst repetition rate to multi-millijoule energies by directly controlling individual intraburst pulse phases, boosting the extractable energy of a chirped pulse amplifier in burst-mode. Control over intraburst phase drift effects is implemented via an active feedback-control loop that is conceptually similar to CEP stabilization developed earlier for continuous pulse trains[23] but is easier to apply because of a simple reference based on linear spectral interference. The developed laser source is shown to be an efficient driver for nonlinear optical applications such as tunable narrowband THz generation via optical rectification. Programmable intense THz-pulse bursts can be applied to efficiently drive compact electron accelerators[28], enhance tailored nonlinear response of molecular gases via rotational stimulation[29], enable scanning-frequency THz spectroscopy, as well as to be used in arbitrary burst applications that are sensitive to the fundamental-frequency pulse envelope but not to its phase. Further, we note that our method is suitable for various CPA layouts and gain materials. While this work is focused on suppressing spectral modes to extract more energy for a certain temporal burst shape, by turning off phase-scrambling and retaining intraburst phase stabilization, we directly obtain a highly promising source of near-infrared frequency combs[30] that offer interesting opportunities for high-acquisition-rate comb spectroscopies.

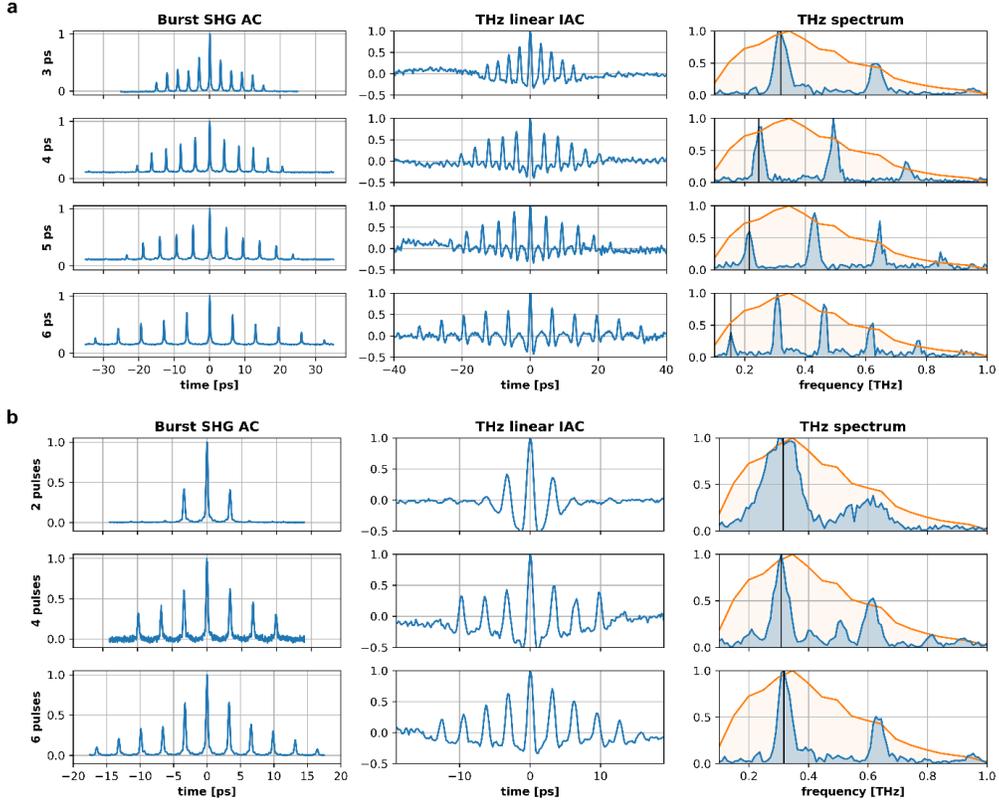

**Fig. 5: Demonstration of control over THz intraburst repetition rates by generation of continuously-tunable narrowband THz-pulse bursts.** Bursts of THz pulses are generated by driving a LiNbO$_3$ crystal with the NIR burst. Second-harmonic autocorrelations of the NIR bursts (left) and linear interferometric autocorrelations from the THz-pulse bursts (middle) were measured. Harmonic THz spectra (right) were retrieved from the latter by Fourier transformation. a) Continuously-tunable THz frequencies can be seen when varying the intraburst pulse spacing with a constant number of N=6 pulses. b) Controllable bandwidths of the THz peaks can be seen when varying the number of pulses with a constant intraburst pulse spacing Δt of 3 ps. Black vertical lines in the THz spectra indicate the intraburst repetition rates.